\documentclass[prb,aps,twocolumn,superscriptaddress,showpacs]{revtex4}
\usepackage{graphicx}
\usepackage{epsfig}
\usepackage{bm}
\begin{document}

\title{Frustrated three-leg spin tubes: from spin 1/2 with chirality to spin 3/2}
\author{J.-B.~Fouet}
\affiliation{Institut Romand de Recherche Num\'erique en Physique des
Mat\'eriaux (IRRMA), CH-1015 Lausanne, Switzerland} 
\author{A.~L\"auchli}
\affiliation{Institut Romand de Recherche Num\'erique en Physique des 
Mat\'eriaux (IRRMA), CH-1015 Lausanne, Switzerland}
\author{S.~Pilgram}
\affiliation{Theoretische Physik, ETH Z\"urich, CH-8098 Z\"urich, Switzerland}
\author{R.~M.~Noack}
\affiliation{Fachbereich Physik, Philipps Universit\"at Marburg, D-35032 Marburg, Germany}
\author{F.~Mila}
\affiliation{
Institute of Theoretical Physics, 
Ecole Polytechnique F\'ed\'erale de Lausanne, 
CH-1015 Lausanne, Switzerland}
\date{\today}

\begin{abstract}
Motivated by the recent discovery of the spin tube
[(CuCl$_2$tachH)$_3$Cl]Cl$_2$, we investigate the properties of a frustrated
three-leg spin tube with antiferromagnetic intra-ring and inter-ring
couplings. We pay special attention to the evolution of the properties from
weak to strong inter-ring coupling and show on the basis of extensive density
matrix renormalization group and exact diagonalization
calculations  
that the system
undergoes a first-order phase transition between a dimerized gapped phase at
weak coupling that can be described by the usual spin-chirality model and a
gapless critical phase at strong coupling that can be described by an
effective spin-3/2 model. We also show that there is a magnetization plateau
at 1/3 in the whole gapped phase and slightly beyond. The implications for
[(CuCl$_2$tachH)$_3$Cl]Cl$_2$ are discussed, with the conclusion that this
system behaves essentially as a spin-3/2 chain.

\end{abstract}
\maketitle

\section{Introduction}

During the last fifteen years, spin ladders have attracted a lot of attention
both from theoretical and experimental physicists (for an early review, see 
Ref.~[\onlinecite{Dagotto1996}]). On one hand, they provide a calculationally tractable
system interpolating between one-dimensional and two-dimensional quantum spin
physics. On the other hand, existing spin ladder compounds allow for a
detailed comparison of theoretical predictions and experimental results. The
interpolation between one and two dimensions is however not smooth: while
half integer spin ladders with an even number of legs $N$ show a gapped spectrum for spin
excitations, ladders with odd $N$ are believed to be gapless (like the
spin-1/2 Heisenberg chain).

For odd $N$-ladders with half-integer spins, applying periodic boundary conditions in the transverse
direction of the ladder (forming thus a {\it spin tube}) yields an even more
intriguing situation: In this case, for antiferromagnetic couplings, the
ground state of an isolated ring is fourfold degenerate (twofold in spin- and
twofold in chirality-space) and the low-energy physics of weakly coupled rings
therefore involves both spin  and chirality degrees of freedom. A number of
theoretical studies have investigated effective low-energy Hamiltonians for
weakly coupled rings by means of bosonization~\cite{Schulz1996,Cabra1998},
density matrix renormalization group (DMRG)~\cite{Kawano1997,L04}, mean field
theory~\cite{Wang01}, and exact diagonalization (ED)~\cite{L04}. All of them
conclude that the ground state is dimerized, displaying a gap for both spin
and chirality excitations. The details of the gaps depend however on the
frustration of the inter-ring coupling~\cite{L04}.

Two experimental candidates for spin tubes are currently available: The
vanadium oxide Na$_2$V$_3$O$_7$~\cite{Millet99,Gavilano03,Gavilano05},
probably a nine-leg spin tube but with only threefold rotational symmetry, and
the three-leg compound [(CuCl$_2$tachH)$_3$Cl]Cl$_2$~\cite{Seeber04,S04}. None
of these examples is in the regime of weak coupling between the rings, intra-
and inter-ring coupling constants being of the same order. It is therefore a
very important first question to ask whether the low-energy physics of these
realistic spin tubes may still be described by the above mentioned
spin-chirality models. In both experimental realizations, neighboring rings
are not coupled by one single antiferromagnetic bond, but by at least two
competing bonds. This leads to additional frustration and raises the second
question 
as to 
how much this frustration 
can change the physics.

We address both questions by 
considering the example of a three-leg tube with
antiferromagnetic intra-ring couplings and two frustrating antiferromagnetic
inter-ring couplings which forms a minimal setup including chirality and
additional frustration.  This model represents directly the compound
[(CuCl$_2$tachH)$_3$Cl]Cl$_2$, but the main results
are expected to apply to other frustrated spin tubes as well.
Regarding the properties, we will concentrate on the presence or 
absence 
of a spin gap, on the nature of the low-lying excitations, and on
the magnetization curve.

The paper is organized as follows: In Sec.~\ref{The model} we introduce the
basic notations and describe several limiting cases of the three-leg spin
tube.  In Sec.~\ref{Numerical results} we discuss the zero-field phase diagram
from a numerical point of view and report on low-spin boundary excitations. In
Sec.~\ref{Effective Hamiltonian Approach} we use continuous unitary
transformations to derive effective low-energy Hamiltonians. 
Finally, we analyze the phase diagram as
a function of the external magnetic field in
Sec.~\ref{Phase diagram under field} and discuss the experimental
implications of our results in Sec.~\ref{compound}.

\section{The model}
\label{The model}
\begin{figure}
  \includegraphics[width=\linewidth]{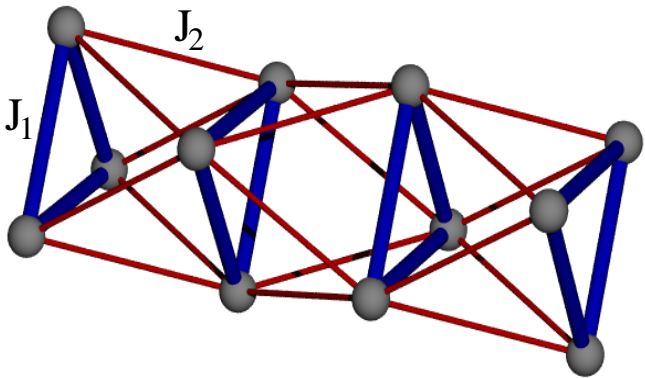}
  \caption{\label{fig-reseau}(Color on-line) Three-leg spin tube with ring
    coupling $J_1$(thick lines) and cross coupling $J_2$ (thin lines).  Both
    couplings are antiferromagnetic, leading to frustration both {\it inside}
    the triangles and {\it between} the triangles.}
\end{figure}

The starting point is the Hamiltonian of the frustrated antiferromagnetic
spin-1/2 Heisenberg model on a three-leg spin tube:
\begin{equation}
H=J_1 \sum_{ij}{\bf \sigma}_{ij}\cdot{\bf \sigma}_{ij+1}
+
J_2 \sum_{ij}{\bf \sigma}_{ij}\cdot\left({\bf \sigma}_{i+1j+1} + 
{\bf \sigma}_{i+1j-1}\right). 
\label{eq-ham}
\end{equation}
The index $i$ runs along the tube, $j=1\dots 3$ around the tube, and ${\bf \sigma}$ is the spin operator for a spin 1/2.  The tube can
be viewed as series of triangles (see Fig. \ref{fig-reseau}).  Each corner of
such a triangle is coupled to two corners on the previous and the next
triangle. In the absence of the $J_1$ coupling, the Hamiltonian is bipartite
(see Fig. \ref{fig-reseau}), and can be seen as a particular wrapping of a
square lattice onto the tube.

\subsection{Lieb Schultz Mattis Theorem}
 \label{LSM}
The Lieb Schultz Mattis (LSM) theorem~\cite{LSM61} is one of the few exact results on 
frustrated spin systems. 
Since its proof is quite general, it can be applied to a wide variety of 
systems with short-range 
interactions~\cite{A88,OA97}, in particular, to the spin tube Hamiltonian~(\ref{eq-ham}). 
An explicit proof of the LSM theorem for this case can be found in Ref.~[\onlinecite{Rojo96}]. 
The theorem states that half-integer spin chains have either a degenerate ground state 
or a gapless spectrum due to a zero-energy mode with momentum $\pi$ relative to the 
ground state. It is interesting to note that both situations are realized in the 
three-leg spin tube system. We first discuss them in two limiting cases. 

\subsection{Weakly coupled triangles}
\label{weakly coupled triangles}
In the limit $J_1 \gg J_2$, the system consists of weakly interacting
triangles. Each isolated triangle has a fourfold degenerate spin-1/2 ground
state that can be labeled by its chirality, $\tau_z$, and its magnetic moment
along the z axis, $S_z$. The projection of the full Hamiltonian~(\ref{eq-ham})
onto the low-energy space spanned by the spin-1/2 states leads to lowest order
in $J_2$ to an effective spin-chirality model,
\begin{equation}
  \begin{array}{rcl}
    H_{1/2} &=&P_{S=1/2}H P_{S=1/2}\\
    &&\\
    &\approx&
    \frac{2J_2}{3}\sum_i
    {\bf S}_i\cdot{\bf S}_{i+1}
    \left(
      1
      +
      \alpha\left[
        {\bf \tau}_i^+\cdot{\bf \tau}_{i+1}^-
        +
        \mbox{h.c.}
      \right]
    \right) \; , 
  \end{array}
  \label{Reduced Chirality}
\end{equation}
where $\alpha=2$ for the frustrated tube of Eq.~(\ref{eq-ham}).  Here, ${\bf
S}_i$ is a spin-1/2 operator describing the total spin on triangle $i$, 
${\bf \tau}_i$ is a pseudo-spin-1/2 operator acting in chirality space, and 
$P_{S=1/2}$ is the projector in the subspace where each triangle is in
a $S=1/2$ state. It can be written as the product of the local
projectors, $P_{S=1/2}=\Pi_i P^{i}_{S=1/2}$, with:
\begin{equation}
  P^i_{S=1/2}=\frac{15-4{\bf S}^2_i}{12}.
  \label{Projector}
\end{equation}
The reduced spin-chirality
Hamiltonian\ (\ref{Reduced Chirality}) has been studied in
Refs.\ [\onlinecite{Kawano1997,L04}]. For $\alpha=2$, it displays a dimerized ground state with a
twofold degeneracy and gaps for both spin and chirality excitations.

\subsection{Strongly coupled triangles}
\label{strongly coupled triangles}
In the limit $J_1=0$, the lattice is effectively bipartite, and we expect
ferromagnetic correlations between spins on the same ring because they belong to the same sublattice. The total spin on each ring should  therefore be close to its maximum value of 3/2, 
and   the physics for $J_1=0$ should be similar to the case of a ferromagnetic
coupling, $J_1 < 0$, within each triangle. The effective Hamiltonian in the
strong coupling case $-J_1\gg J_2$ reads:
\begin{equation}
  H_{3/2} =
  K_1 \sum_{i} {\bf S}_i\cdot{\bf S}_{i+1}
  +
  K_2 \sum_{i} \left({\bf S}_i\cdot{\bf S}_{i+1}\right)^2.
  \label{Reduced Quadruplet}
\end{equation}
where ${\bf S}_i$ is now a spin-3/2 operator.  The coupling constants are
given by $K_1 = 2J_2 / 3 + J_2^2 / (18|J_1|)$ and $K_2 = - J_2^2 / (54|J_1|)$.
Note that the nature of the second order corrections to $K_1$ and $K_2$ does
not seem to frustrate the effective spin-3/2 chain.  It is therefore
reasonable to argue that the spin-3/2 state is stabilized at least up to
$J_1=0$ even though this value lies beyond the perturbative limit.  This
argument can be made more quantitative by performing range-two
CORE-calculations\cite{MW96,CLM04} yielding effective coupling constants $K_1
\approx 0.78$ and $K_2 \approx -0.04$ for $J_1=0$ (the bicubic $K_3$ coefficient
is 4 times smaller than $K_2$). The ratio of the couplings does not change 
drastically even by increasing $J_1/J_2$ up to $1$. The physics of the effective
Hamiltonian\ (\ref{Reduced Quadruplet}) is therefore dominated by the Heisenberg ($K_1$)
term, which is known to belong to the universality class of a Luttinger liquid
\cite{HWHM96}, and therefore has a gapless spectrum. We believe that small biquadratic
or bicubic interactions will not immediately open a gap \cite{ZimanSchulz} and
we expect the low energy theory to remain stable.
Note that these gapless excitations are in sharp contrast to the gapped spectrum 
of the spin-chirality Hamiltonian\ (\ref{Reduced Chirality}).

Based on the preliminary arguments of subsections 
\ref{weakly coupled triangles} and \ref{strongly coupled triangles}, we expect the
frustrated spin-tube Hamiltonian\ (\ref{eq-ham}) to undergo at least
one quantum phase 
transition as we move from weakly coupled ($J_1 \gg J_2$) to strongly coupled
($J_1 \ll J_2$) triangles.

\section{Numerical results}
\label{Numerical results}

In the following, we investigate the properties of this phase transition
numerically. We have performed extensive density matrix renormalization group
(DMRG) calculations\cite{SW92} on tubes with open boundaries and $L\le100$
and exact diagonalization (ED) on small systems up to $L=12$ using periodic
boundary conditions (PBC).

Two independent quantities prove clearly that there is a single first-order
bulk phase transition. They are discussed in Secs.~\ref{local spin on a ring}
and \ref{dimerization} respectively. Furthermore, on open boundary systems,
there is a second ``transition'' at which the state of rings at the boundary
jump from predominantly $S=1/2$ to predominantly $S=3/2$. This phenomenon will
be discussed in Sec.~\ref{boundary excitation}.

\subsection{Local spin on a ring}
\label{local spin on a ring}
\begin{figure}
  \includegraphics*[width=\linewidth]{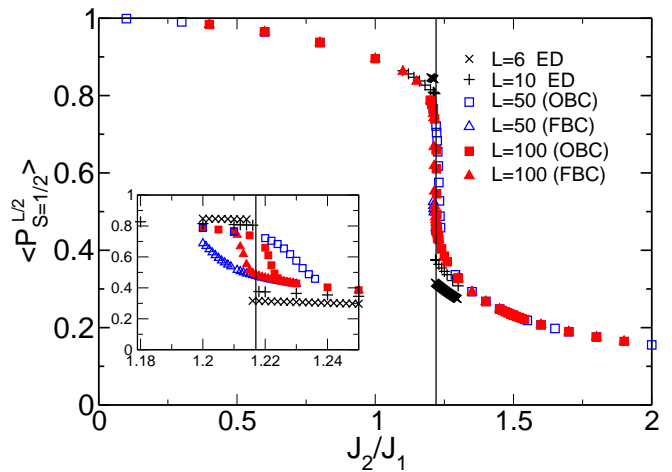}
  \caption{(Color on-line) Expectation value of the projector
    $P^{L/2}_{S=1/2}$ in the bulk as a function of $J_2/J_1$ for different
    lengths and boundary conditions (see text). The jump at $J_2/J_1 \approx
    1.22$ is a clear indication of a first-order transition.
    \label{PS1half-bulk}}
\end{figure}

The first-order character is best seen in the ground state expectation value
$\langle P^i_{S=1/2} \rangle$ of the previously defined
projector~(\ref{Projector}), which is a purely local quantity defined on a
single triangle $i$~\cite{W02}.  We have computed this expectation value as a function of
$J_2$. The results are shown in Fig.\ \ref{PS1half-bulk}, which displays the
projector in the middle of the tube at $i=L/2$. For weak coupling, $J_2/J_1\ll
1$, the expectation value is close to one, for strong coupling, $J_2/J_1\gg
1$, it is close to zero. The expectation value clearly jumps at a value
$J_{2c}=(1.219\pm 0.003) J_1$ indicating a first-order phase transition which is
magnified in the inset of Fig.\ \ref{PS1half-bulk}.

\begin{figure}
  \includegraphics*[width=\linewidth]{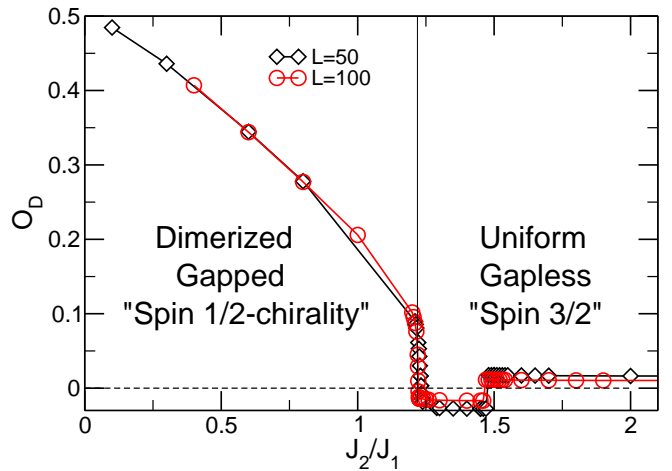}
  \caption{(Color on-line) Dimerization at the center of the tube $O_D$ as a
    function of $J_2/J_1$ for $L=50$ (black squares) and $L=100$ (red
    circles).  The vertical line represents the disappearance of the true
    order parameter: The dimerization above this line vanishes after proper
    finite-size scaling.
    \label{fig-dim2}}
\end{figure} 

Several comments are in order: Fig.\ \ref{PS1half-bulk} displays results
obtained both with DMRG and ED. 
To check the importance of boundary effects 
(see Sec.\ \ref{boundary excitation}), DMRG was performed with two types 
of boundary conditions: the
standard Open Boundary Conditions (OBC), and the Ferromagnetic Boundary
Conditions (FBC), where we put a ferromagnetic coupling $J_1=-10$ on the two
triangles at the boundary of the tube, thereby strongly favoring the $S=3/2$ state.
For the ED results, periodic boundary conditions were applied. The curves for
the three kinds of boundary conditions differ only near the first-order
transition, and the differences decrease with the increasing length of the
tube.  In the ED, we additionally observe that for system sizes $L=4p+2$ the
ground state changes the momentum sector at the transition from $k=0$ for weak
coupling to $k=\pi$ for strong coupling. This is another fingerprint of a
first-order transition.

\begin{figure}
  \includegraphics*[width=\linewidth]{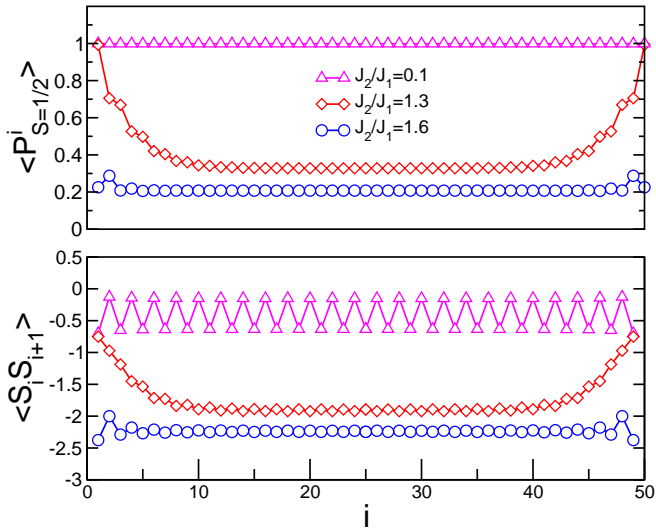}
  \caption{\label{fig-PS+dimvsx} (Color on-line) Spatial dependence of the
    projector $\langle P^{i}_{S=1/2} \rangle$ (upper panel) and
    nearest-neighbor correlation function $\langle {\bf S}_i\cdot\bf{S}_{i+1}
    \rangle$ (lower panel) for $L=50$ and different values of the inter-ring
    coupling $J_2$.  } 
\end{figure}

\subsection{Dimerization}
\label{dimerization}
The first-order character of the phase transition is also clearly apparent in
inter-ring correlations. As an example, we investigate the local dimerization
defined as
$$ 
D_i=(-1)^i\left({\bf S}_i.{\bf S}_{i+1}-{\bf S}_{i+1}.{\bf S}_{i+2}\right).
$$ 
This dimerization can be viewed as the order parameter of the
symmetry-broken weak coupling phase.  It is also present in the spin-chirality
model discussed in Sec.\ \ref{weakly coupled triangles} for which the
dimerization opens a spin gap~\cite{Kawano1997,Wang01,L04}. Since we work with
open boundary conditions, the quantity $D_i$ varies with the
ring position $i$. Fig.\ \ref{fig-dim2} shows the order parameter
$O_D(L)=\langle D_{L/2}\rangle$ (the dimerization in the middle of the tube)
for different system sizes $L$.  Two sharp transitions are observed, the first
one around $J_2 =J_{2c}$ and the second one at a higher value $J_2/J_1\approx
1.47$. However, finite-size scaling shows that the order parameter $O_D(L)$
remains finite only for $J_2 <J_{2c}$ in the thermodynamic limit. The first
transition, the disappearance of the dimerization, corresponds to the phase
transition to a $S=3/2$ phase on the rings. The second transition is a
boundary effect discussed in the next section.

\subsection{Boundary excitations}
\label{boundary excitation}
Boundary spin-1/2 degrees of freedom in open spin $S=3/2$ chains
have been predicted theoretically in Ref.~[\onlinecite{N94}], and were later
confirmed in DMRG studies of unfrustrated\cite{QNS95} and 
frustrated\cite{R98} spin $S=3/2$ chains.

In our model we find these edge states as well for sufficiently large
$J_2/J_1$. However approaching the first-order transition coming from
$J_2>J_{2c}$, there is a second class of edge states, which are related 
to a kind of nucleation of the dimerized $S=1/2$ phase at the boundaries.
These edge states are different than those discussed above, as they now 
originate from the $S=1/2$ subspace of a ring, and therefore also include a
chirality degree of freedom.

This is most clearly seen in the spatial dependence of the correlation
functions considered in the two subsections above. The upper panel of Fig.~\ref{fig-PS+dimvsx} shows the space
dependence of the projector $\langle P^i_{S=1/2} \rangle$ and the lower panel
of the nearest-neighbor spin correlation $\langle {\bf S}_{i} \cdot {\bf
S}_{i+1}\rangle$ for different values of $J_2/J_1$.  The ring at the boundary stays in a predominant $S=1/2$
state even for $J_2>J_{2c}$, where the bulk transition has already
occurred. The boundary spin-1/2 states disappear only beyond $J_2 \approx 1.5$.
The nearest-neighbor spin-spin correlation function follows a similar behavior:
the correlations at the boundary are close to those obtained in the bulk
$S=1/2$ phase, and then gradually approach the values of the bulk $S=3/2$
phase.

\begin{figure}
  \includegraphics[width=\linewidth]{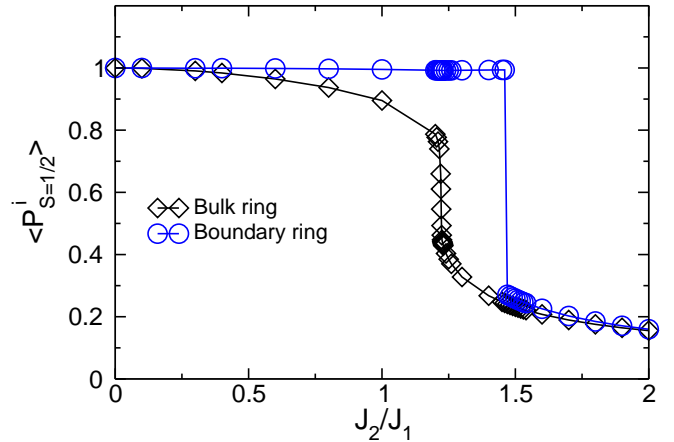}
  \caption{\label{fig-boudaryN300} (Color on-line) The projector $\langle
    P^{i}_{S=1/2} \rangle$ in the bulk and at the boundary of the tube for
    $L=100$ jumps for different critical couplings indicating the existence of
    a boundary excitation.}
\end{figure}

The different behavior of bulk and edge rings are nicely illustrated in
Fig.~\ref{fig-boudaryN300}, in which the values of $\langle P_{(S=1/2)}\rangle$
in the bulk and at the boundary of the tube are plotted as a function of
the coupling ratio $J_2/J_1$. The expectation value of the projector drops
at $J_2/J_1 \approx 1.22$, whereas the boundary ring remains
in a $S=1/2$ state up to $J_2/J_1 \approx 1.5$.

\section{Effective Hamiltonian Approach}
\label{Effective Hamiltonian Approach}

Since the numerical results strongly indicate that the transition is
first order, it should be possible to describe the transition as occuring directly
between an effective spin-chirality model and a spin 3/2 model.
Indeed, we have previously discussed that the
ground state of the model Hamiltonian~(\ref{eq-ham}) lies in the spin-1/2
sector for $J_1 \gg J_2$ and in the spin-3/2 sector for $J_2 \ll J_1$. The
problem is then reduced to diagonalizing an effective
Hamiltonian~(\ref{Reduced Chirality}) or (\ref{Reduced Quadruplet}) on a
truncated Hilbert space (for the spin-1/2 case see Refs.\
[\onlinecite{Kawano1997,L04}]).  However, in the intermediate regime, $J_1 \sim J_2$,
it is no longer possible to construct effective Hamiltonians using
perturbation theory in the small parameter $J_2 / \left|J_1\right|$ and an
alternative construction has to be found.  In this context, it is interesting
to observe that the above Hamiltonian $H$ can be extended to
$$
H_e
=
H +
J_2' \sum_{ij}
{\bf \sigma}_{ij}
\cdot
{\bf \sigma}_{i+1j}.
$$
For $J_2=J_2'$, the extended Hamiltonian $H_e$ has the special property to
conserve total spin and chirality of each triangle
separately~\cite{Honecker2000}. Note that this property is independent of the
relative strength of $J_1$ and $J_2$ and holds also for $J_1 \sim J_2$.  The
Hamiltonian $H_e$ is naturally block-diagonal when  $J_2=J_2'$, each block being
characterized by a set of quantum numbers for spin and chirality on each
triangle. The ground state is found either in the block where all triangles
have spin-1/2 or where all have spin-3/2. Both cases are obviously connected
by a first order phase transition.

\begin{figure}
  \includegraphics*[width=\linewidth]{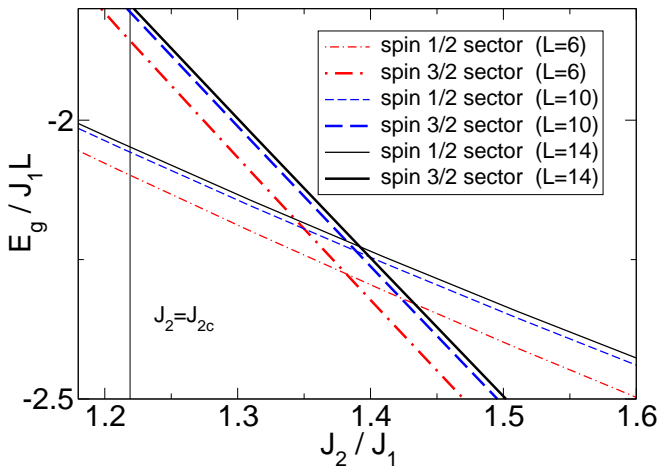}
  \caption{ (Color on-line) Comparison of ground state energies in the
    spin-1/2 and spin-3/2 sectors as a function of $J_2/J_1$ with 15 couplings kept in the flow. Finite-size
    effects only appear for short tubes of length $N=6,8$. The first-order
    phase transition is located near $J_2 = 1.4J_1$. The vertical line represents the position of the phase transition
according to the numerical results.  The discrepancy between the two is due to the truncation in the space of possible
    couplings.  \label{Ground State Energies}}
\end{figure}

For our Hamiltonian $H$ we have $J_2\ne J_2' =0$ and the large set of
conserved quantum numbers is lost. Even the total number of spin-1/2 triangles
is no longer conserved. To restore this latter conservation law, we propose to
apply a continuous unitary transformation to the Hamiltonian $H$ which
eliminates all couplings that change the total number of spin-1/2
triangles. The obtained effective Hamiltonian $H_{\text{eff}}$ is again
block-diagonal and allows us to calculate the ground state energies of the
spin-1/2 and spin-3/2 sector independently. We may then infer the critical
$J_{2c}$ for the first order phase transition from the comparison of those
energies. The proposed unitary transformation can be carried out exactly in
the limit $J_2 - J_2'\rightarrow 0$. In our case, $J_2'=0$, the construction
of the effective Hamiltonian $H_{\text{eff}}$ can be viewed as a perturbative
expansion in the parameter $J_2 - J_2'$.

A powerful tool for the derivation of effective Hamiltonians was introduced by
Wegner~\cite{Wegner1994}, the so-called flow equations which were first
applied to spin Hamiltonians by Uhrig~\cite{Uhrig1998}. Flow equations apply
infinitesimal unitary transformations (parametrized by $\ell$) to the original
Hamiltonian until an effective Hamiltonian with the desired properties is
obtained
\begin{equation}
\frac{dH(\ell)}{d\ell} = \left[\eta(\ell),H(\ell)\right],
\qquad
H(\ell \rightarrow \infty) \rightarrow H_{\text{eff}}
.
\label{Flow Equation}
\end{equation}
Following Ref.\ [\onlinecite{Wegner1994}], we choose the commutator
\begin{equation}
\eta(\ell) = \left[
H_c(\ell),H(\ell)
\right]
\label{Generator}
\end{equation} 
as an appropriate generator of the transformation.  Here, $H_c(\ell)$
denotes the conserving part of the Hamiltonian which does not change the
number of spin-1/2 triangles.  The above defined flow generates an infinite
set of couplings which has to be suitably truncated. To illustrate the
behavior of the flow qualitatively, we first use a truncated set of only
three couplings. Later on, we will use a set of sixteen couplings which allows
us to arrive at semi-quantitative results.

For the qualitative analysis, we decompose the Hamiltonian 
$$
H(\ell) =  J_1(\ell)h_1 + J_{2,c}(\ell) h_{2,c}
+ J_{2,n}(\ell) h_{2,n},
$$ 
into the coupling
$$
h_1 = \sum_{ij}{\bf \sigma}_{ij}\cdot{\bf \sigma}_{ij+1},
$$
the coupling $h_{2,c}$ which is the part of
$$
h_2 = \sum_{ij} 
{\bf \sigma}_{ij}\cdot\left({\bf \sigma}_{i+1j+1} + 
{\bf \sigma}_{i+1j-1}\right)
$$ that conserves the number of spin-1/2 triangles and the coupling $h_{2,n}=
h_2 - h_{2,c}$ that changes the number of spin-1/2 triangles. The two
couplings $h_{2,c}$ and $h_{2,n}$ are readily obtained by applying the
projectors\ (\ref{Projector}) to the terms of $h_2$.

Inserting the Hamiltonian $H(\ell)$ and its conserving part $H_c(\ell)=
 J_1(\ell)h_1 + J_{2,c}(\ell) h_{2,c}$ into
Eqs.\ (\ref{Flow Equation}) and (\ref{Generator}), we obtain (after the
truncation) three coupled differential equations for the coupling constants
$$
\frac{dJ_1}{d\ell} =
\left(32J_1 - 21\frac{1}{3}J_{2,c}\right)(J_{2,n})^2
,
$$
$$
\frac{dJ_{2,c}}{d\ell} =
\left(
-13\frac{5}{7}J_1 + 19\frac{19}{63}J_{2,c}
\right)
(J_{2,n})^2
,
$$
\begin{equation}
\label{Simple Flow}
\frac{dJ_{2,n}}{d\ell} =
\left(
-72 J_1^2
+96 J_1J_{2,c}
-67\frac{5}{9}J_{2,c}^2
\right)
J_{2,n}
\; ,
\end{equation}
with the initial conditions
$$
J_1(0) = J_1, J_{2,c}(0)=J_{2,n}(0)=J_2.
$$

The analysis of the last differential equation shows that $J_{2,n}$ scales to
zero during the flow, leading to an effective Hamiltonian $H_{\text{eff}}$
which obeys the desired conservation law.  The qualitative inspection of the
first and second equation shows that for $J_2 < 1.5J_1$ the spin-1/2 phase is
stabilized ($J_1(\infty) > J_1(0)$, $J_{2,c}(\infty) < J_{2,c}(0)$), whereas
for $J_2 > 1.5J_1$ the spin-3/2 phase is stabilized in the same way.  For very
large initial $J_2$ the effective $J_1(\infty)$ becomes even ferromagnetic
$J_1(\infty)<0$, so that the one-site coupling $h_1$ itself favors the
spin-3/2 state. This analysis is confirmed by the numerical integration of
Eqs.\ (\ref{Simple Flow}).

The accuracy of the above reasoning can be made semi-quantitative, when more
couplings are included in the flow. For this purpose we evolve the
one-triangle coupling $h_1$ and all the fifteen two-triangle nearest-neighbor
couplings which are allowed by symmetry under the flow. We  evaluate the ground
state energies in the spin-1/2 and spin-3/2 sectors of $H_{\text{eff}}$ by exact diagonalization
of small chains corresponding to tubes up to length $L=14$. (Note that the dimension of the effective
Hilbert space is reduced in comparison to the full Hilbert space by a factor
of $2^L$).  The results of the numerical calculation are shown in Fig\
\ref{Ground State Energies}. The phase transition between a spin-1/2-chirality
chain and a spin-3/2 chain appears around $J_2 \approx 1.4J_1$. This value
lies slightly above the precise $J_{2c}$ obtained by DMRG. As can be seen from
Fig.\ \ref{Ground State Energies}, finite-size effects do not explain this
quantitative discrepancy.  It is in fact due to the truncation of the flow
equations to nearest-neighbor couplings.
It can be checked numerically that the effect of including next-nearest
neighbor two-triangle couplings in the flow is negligible. To reach a more
quantitative result, it is necessary to include three-triangle couplings.

In summary, a first-order transition between a spin-chirality model and a
spin-3/2 model is also found in this approach, and the agreement with the
critical ratio $J_2/J_1$ found in DMRG is very satisfactory considering the
truncation of the flow. Therefore we believe that the case for a first-order
transition is very strong.

\section{Phase diagram in a field} 
\label{Phase diagram under field}

Finally, to make contact with the experiments performed on 
$[({\rm CuCl}_2{\rm tachH})_3{\rm Cl}]{\rm Cl}_2$,
we investigate the properties of the model in the presence of
an external magnetic field.

\begin{figure}
  \includegraphics[width=\linewidth]{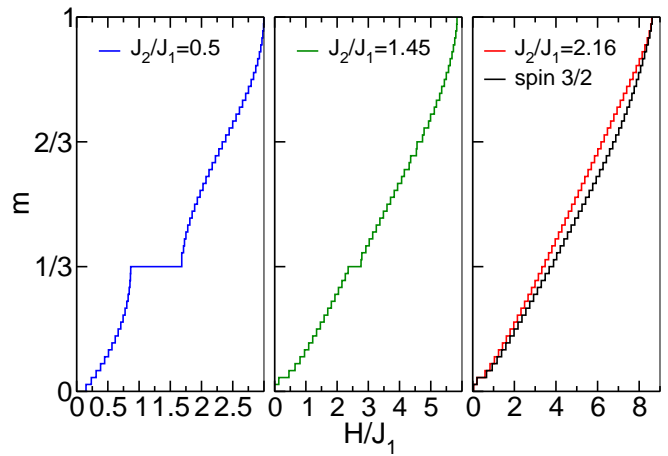}
  \caption{
    \label{mag-exp} (Color on-line) 
    Magnetization curves for a spin tube with $L=36$ and three different
    values of $J_2/J_1$. In the right panel, we also show the magnetization curve
    of a $S=3/2$ chain for comparison.}
\end{figure}

For $J_2/J_1=0$, the triangles are decoupled and we expect a magnetization
jump from $m=0$ to $m=1/3$ at $h=0^+$, where all the triangles are in a spin-1/2 state with $S_z=+1/2$.  This $m=1/3$ plateau should survive in a finite
parameter range. 
In terms of the spin-chirality model of Ref.~[\onlinecite{L04}],
such a plateau is quite unusual. Indeed, when the spin degrees of freedom align
ferromagnetically in the field, the effective model for the chirality becomes
an XY model, whose spectrum is gapless. So, according to this argument, this
plateau phase is expected to have a gap to spin excitations -- as do 
all plateau 
phases -- 
and to simultaneously 
possess 
gapless chirality excitations. Note
that higher-order terms in the effective model might open a small gap in the
chirality excitations as well, but being a higher order effect, this gap (if
any) should be small compared to the spin gap, and should thus lead to
observable properties, such as a specific heat linear in $T$ below the
temperature scale set by the width of the plateau.

\begin{figure}
  \includegraphics[width=\linewidth]{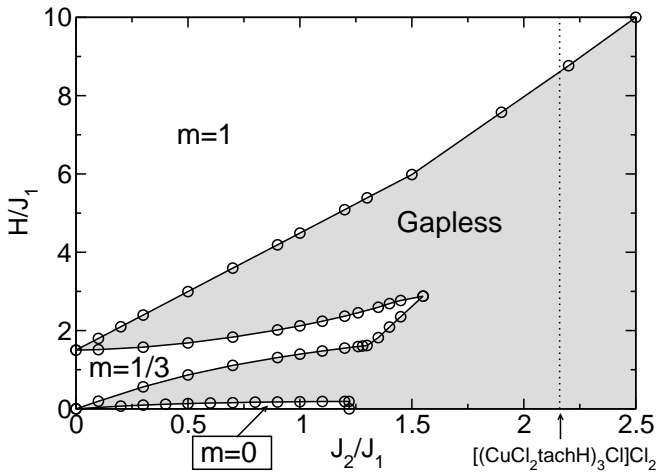}
  \caption{\label{phase-diagram} Phase diagram as a function of $J_2/J_1$ and
    of the magnetic field $H$.  Symbols are numerical data and solid lines are
    guides to the eye for the boundaries of the phase transitions.  The gray
    area is a gapless phase, and the white area are plateau phases with a spin
    gap.  Dotted line: experimental value of $J_2/J_1$ for the compound of
    Ref.\protect{[\onlinecite{S04}]}}
\end{figure}

Typical magnetization curves are shown in Fig.~\ref{mag-exp} for several
values of $J_2/J_1$. As long as the ratio $J_2/J_1$ is not too large, a
plateau at $m=1/3$ is clearly present.
As can be seen in Fig. \ref{phase-diagram}, this plateau phase survives
everywhere in the region 
in which the system is effectively a spin-1/2 chain with
chirality, but interestingly, 
it extends beyond that point up to
$J_2/J_1\approx 1.6$. 
It is plausible that the region of validity of the effective spin-1/2-chirality model extends to higher values of $J_2/J_1$ when a magnetic field
is applied to the system. Indeed, an abrupt jump in $P^{L/2}_{S=1/2}$
is still present for slightly polarized samples
and the corresponding critical value of the ratio $J_2/J_1$ increases slightly 
with magnetization.
By the time the $m=1/3$ plateau is reached, the suppression of $P^{L/2}_{S=1/2}$
is very smooth, however, and it is not possible to decide on the basis
of the available data whether the first-order line extends up to the
right boundary of the plateau phase, or whether it ends at a critical point
on the way.
In any case, for large $J_2/J_1$, the magnetization curve becomes
smooth, as it should for a spin-3/2 chain.

\section{$[({\rm CuCl}_2{\rm tachH})_3{\rm Cl}]{\rm Cl}_2$}
\label{compound}

\begin{figure}
  \includegraphics*[width=\linewidth]{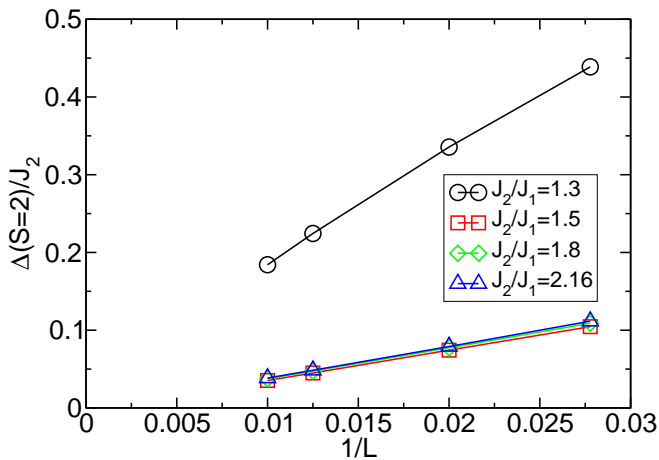}
  \caption{
    \label{gap-exp}
    (Color on-line) Gap to the S=2 sector  as a
    function of 1/L for several values of $J_2/J_1$}
\end{figure}

The Hamiltonian of Eq.~(\ref{eq-ham}) is believed to be realized in the
compound [(CuCl$_2$tachH)$_3$Cl]Cl$_2$, with $J_2/J_1 \approx 2.16$, and the
theoretical investigation of that model reported in Ref.~[\onlinecite{S04}] came to
the somewhat surprising conclusion that the system has a gap to all (spin and
singlet) excitations. This is at odds with the LSM theorem (see section \ref{LSM}).
  On the basis of our DMRG results, we do not reach the
same conclusion.  If $J_2/J_1$ is small enough, the system should be in a
dimerized phase with a spin gap and a twofold degenerate ground state. Hence a
low-lying singlet excitation should be present in finite systems and should
collapse onto the ground state in the thermodynamic limit.  If $J_2/J_1$ is
large enough, the system should essentially behave as a spin-3/2 chain and be
gapless with low-lying spinon excitations. The phase transition between these
phases takes place for $J_2/J_1\approx 1.22$. So, with a ratio $J_2/J_1
\approx 2.16$, as deduced from the temperature dependence of the
susceptibility in Ref.~[\onlinecite{S04}], we predict that
[(CuCl$_2$tachH)$_3$Cl]Cl$_2$ should have a gapless spectrum, and, since this
ratio is also larger than the critical value that marks the disappearance of
the 1/3 plateau, there should be no plateau at 1/3. In fact, according to the
discussion of the previous section, the absence of a 1/3 plateau in the
magnetization curve of [(CuCl$_2$tachH)$_3$Cl]Cl$_2$ reported in
Ref.~[\onlinecite{S04}] shows unambiguously that this compound lies in the effective
$S=3/2$ strong coupling phase.

To be more specific, the claim regarding the absence of a spin gap is clearly
supported by our calculations of the gap to $S=2$ excitations (since we
are working with open boundary conditions, the gap to $S=1$ excitations is very
small and corresponds to boundary excitations\cite{N94,QNS95}). 
Indeed, the finite-size scaling of the gap is consistent with a vanishing value
in the thermodynamic limit for all ratios $J_2/J_1$ larger than 1.22 
(see Fig.~\ref{gap-exp}).

Regarding the magnetization curve, our results agree with those of
Ref.~[\onlinecite{S04}], which were obtained on smaller systems: The curve is smooth
with no trace of any irregularity 
close to 1/3-magnetization.  Fig. \ref{mag-exp}
shows the magnetization curve for the spin tube for various value of $J_2$  and
for the spin-3/2 chain for $L=36$.  The coupling of the spin-3/2 chain has
been chosen so that the saturation field is the same as the saturation field 
for the experimental value of the couplings ($J_2/J_1=2.16)$. The two
magnetization curves are very similar 
giving further support to our claim that [(CuCl$_2$tachH)$_3$Cl]Cl$_2$
should be 
regarded 
as a spin-3/2 chain. It would be desirable to have additional 
low-temperature
susceptibility data in order to clarify whether the compound really shows a spin
gap. On the basis of the available susceptibility and magnetization data, gapless
behavior is clearly not excluded.

\section{Conclusions}

We have investigated three-leg spin tubes consisting of rings that are coupled by
competing antiferromagnetic bonds.  We have shown that this frustration
can drive the system away from the typically used 
effective spin-1/2-chirality
model of regular spin tubes.  It becomes an effective spin-3/2 chain 
even if the inter-ring couplings are antiferromagnetic. In the specific case
of the model relevant to [(CuCl$_2$tachH)$_3$Cl]Cl$_2$ in which each spin of a
ring is coupled to two spins of the neighboring rings, the transition between
these phases as a function of the inter-ring coupling has been shown to be
first order. This should be contrasted to the case of the unfrustrated spin
tube, in which case the transition is expected to be second order and to take
place when 
the inter-ring coupling changes sign. These results lead
to a new interpretation of the properties of [(CuCl$_2$tachH)$_3$Cl]Cl$_2$,
and to specific predictions regarding the occurrence of a plateau at 1/3 in
this geometry. It is our hope that the present work will motivate further
investigations of the excitations of [(CuCl$_2$tachH)$_3$Cl]Cl$_2$ to check
our predictions, and more generally of the (so far) small but fascinating
family of spin tubes.

\section*{Note added}
After submission of this work, K. Okunishi {\it et al.} informed us  that they
 had submitted a work reaching the same conclusion concerning the existence
 of a first order phase transition\cite{Oku2005}.

\section*{Acknowledgments}
We are thankful to Sabrina Rabello, who was involved at an early stage of this
project. This work was supported by the Swiss National Fund and by
MaNEP. Calculations have been partly performed on the Pleiades cluster of
EPFL.  \bibliographystyle{prsty}

\end{document}